\documentclass[12pt]{article}
\usepackage{graphicx}
\begin{document}

\title{Would the Existence of CTCs Allow for Nonlocal Signaling?
}


\author{Lucas Dunlap}



\date{}

\newcommand{\ket}[1]{\left|#1\right>}
\newcommand{\kett}[1]{\(\left|#1\right>\)}
\newcommand{\bra}[1]{\left<#1\right|}
\newcommand{\brat}[1]{\(\left<#1\right|\)}

\maketitle

\begin{abstract}
A recent paper from Brun et al.\ has argued that access to a closed timelike curve (CTC) would allow for the possibility of perfectly distinguishing nonorthogonal quantum states. This result can be used to develop a protocol for instantaneous nonlocal signaling. Several commenters have argued that nonlocal signaling must fail in this and in similar cases, often citing consistency with relativity as the justification. I argue that this objection fails to rule out nonlocal signaling in the presence of a CTC. I argue that the reason these authors are motivated to exclude the prediction of nonlocal signaling is because the No Signaling principle is considered to a fundamental part of the formulation of the quantum information approach. I draw out the relationship between nonlocal signaling, quantum information, and relativity, and argue that the principle theory formulation of quantum mechanics, which is at the foundation of the quantum information approach, is in tension with foundational assumptions of DeutschÕs D-CTC model, on which this protocol is based.

\end{abstract}

\section{Introduction}

Quantum information theory (QIT) has been a majorly productive research program over the last two decades. It has enabled scientists to make progress theoretically, experimentally, and in terms of the development of technology. However, our understanding of what quantum information is telling us about the world---and what it is telling us about quantum theory itself---remains underdeveloped. Broadly speaking, this paper is an attempt to reconcile the conceptual framework at play in the metaphysical approach to the foundations of quantum mechanics with the framework at play in quantum information science.

In this paper, I focus on a debate about the predicted behavior of quantum systems in the presence of a localized region of spacetime subject to nonlinear laws of evolution. The particular example under consideration is that of a closed timelike curve, or CTC. In this literature, the peculiarities of how time travel could possibly be achieved are not addressed. Rather, CTCs are treated as a resource. In this context, we can formulate the question ``what could we do if we had access to a CTC?"

David Deutsch's seminal 1991 paper \cite{Deutsch1991} set the groundwork for a quantum mechanical analysis of the information-processing capabilities of a quantum system augmented by access to a CTC. Over the last decade, interest has flourished in the particular computational tasks that can be achieved with a CTC-assisted quantum computer circuit. However, a debate has arisen surrounding a particularly strange result: the ability to distinguish non-orthogonal quantum states. This is impossible according to ordinary quantum theory, but the Deutsch's analysis of the behavior of quantum systems in the presence of a CTC seems to predict it.\footnote{In this paper, I will be working exclusively with Deutsch's CTC model for quantum systems (D-CTCs). There is an alternative proposal for how to understand the behavior of quantum systems in the presence of CTCs, referred to as P-CTCs. While there are significant differences between the predictions and underlying physics of the two proposals, both allow for the behavior under consideration in this paper, i.e.\ distinguishing non-orthogonal states, and Nonlocal Signaling. In fact, Gisin showed \cite{Gisin1990} that a more general nonlinear framework would allow for the same behavior.} Furthermore, this ability leads to an even more radical conclusion: quantum CTCs allow for information to be sent between arbitrarily distant parties instantaneously.

However, there has been serious resistance to this conclusion from various parties to the debate. And the attempts to formulate exactly why this admittedly non-quantum-mechanical behavior should be ruled out has illuminated the underlying assumptions that are often at play, even in a field such as quantum information, which purports to be formulated entirely in operationalist terms, and neutral with respect to interpretational debates.

A CTC-assisted quantum computational circuit may seem like an exotic example. But analyzing these kinds of systems has proved to be very fruitful. Working with this example has brought to light several common confusions about one of the central concepts of quantum theory: nonlocality. In the foundations literature, quantum nonlocality is often closely connected to the concepts of information and causation. For example, it is often said that the nonlocal correlations of quantum mechanics are allowed at spacelike separation because no information is traveling between the two distant systems \(A\) and \(B\) faster than a light signal could. In cases where nonlocality is exploited as an information channel (as in the quantum teleportation protocol) then quantum information is not present (or useable) at \(B\) until after a classical (lightlike) signal is received from \(A\).

I will argue that the exploitation of quantum correlations to send a message---as seemingly allowed by the existence of CTCs---is not itself inconsistent with relativity. The relativistic constraints that disallow signaling do so because of the potential to send information to the past, giving rise to the possibility of a paradox. I will argue, however, that Deutsch's consistency condition, which constrains the possible nonlinear evolution of systems interacting with closed loops of information on CTCs such that no contradiction can obtain, should apply to this ``radio to the past" as well.

The predicted existence of instantaneous signaling is contrary to a foundational principle of the quantum information-theoretic approach. The No-Signaling Principle plays a fundamental role in the formulation of quantum information. I argue that this accounts for the resistance to the conclusion that it can be violated under certain conditions. Furthermore, I argue that the No-Signaling Principle's inclusion as a fundamental postulate about the nature of the quantum world, as is the case in the quantum information-theoretic interpretation of quantum theory, as advocated in Bub and Pitowsky's ``Two Dogmas About Quantum Mechanics'', represents a commitment to a principle-theoretic conception of quantum theory.

Whereas a ``constructive'' theory is built up from its ontology and dynamics, the fundamental formulation of a ``principle'' theory is ``top-down'', in terms of inviolable global principles (the paradigm case is special relativity). Principle theories explicitly deny the fundamentality of a theory's ontology, and consider constructive formulations of theories to be secondary, and to have a role only as proofs of the consistency of their principles. 

This paper will examine the recent debate surrounding this point in the foundations literature. I will address the relativistic argument for the impossibility of superluminal signaling. That argument fails to apply in the context of this example, and I will argue that there is a deeper motivation at play for ruling out the possibility of signaling, which has to do with the fundamental commitments of the quantum information approach. I argue, however, that the framework developed by Deutsch is inconsistent with these commitments, and therefore the justification for including a No Signaling principle in the D-CTC framework offered by quantum information theorists is indefensible. 


\section{Deutsch's Circuit Model for CTCs}

In his well known 1991 paper \cite{Deutsch1991}, Deutsch introduced a model for the analysis of the physical behavior of CTCs.  Prior to his work, the standard way of analyzing the physical effects of chronology-violating regions of spacetime was in terms of their underlying geometry.  Deutsch considered this approach to be insufficient because it fails to take quantum mechanical effects into account.  He proposed an alternative approach which involves analyzing the behavior of CTCs in terms of their information processing capabilities.  

He begins his account by defining a notion of equivalence between spacetime-bounded networks containing chronology-violating regions.  A network in this context is to be understood as a spacetime geometry which takes as input the initial state of a physical system and outputs the system's final state.  Two networks are \emph{denotationally equivalent} if their outputs are the same function of their inputs.  That  is to say, regardless of whether two networks have differing spacetime geometries, if the function that maps their initial states to their final states is the same, they are denotationally equivalent.

Next he introduces the idea that the transformation between any two denotationally equivalent networks is trivial.  Insofar as we are interested in analyzing CTCs in terms of their physical effects (that is, their output given a certain input), we are free to use the simplest model available in the denotational equivalence class of a particular network for the purpose of our analysis of the information flow through a CTC.

The final step of his proposal is to introduce a simple standard form into which any spacetime-bounded network can be trivially transformed for the purpose of analysis.  The simple standard form involves translating all spacetime-bounded networks into circuits in which each particle traveling in the original network is replaced by sufficiently many carrier particles, each of which have a single 2-state internal degree of freedom (a bit).  The regions in which the particles interact are localized (by denotationally trivial transformations) into gates, such that the states of the particles so not evolve while traveling between them.  And finally, all chronology-violating effects of the network are localized to sufficiently many carrier particles on closed loops, which only interact with chronology-respecting particles in gates.

Deutsch points out that chronology violation itself makes no difference to the behavior of a network unless there is a closed loop of information.  In the original network, this closed information path could potentially not be confined to the trajectory of any single particle (since the carriers can interact with each other), but for any such network, there is a denotationally trivial transformation which will localize the closed loop of information on sufficiently many carriers on closed paths.

The real innovation of this approach is that it can very easily accommodate quantum mechanical effects by relaxing the requirement that the carrier particles be in a well-defined classical state after interactions.  If viewed classically, networks containing chronology violations can lead to paradoxes that seem to put unnaturally strong constraints on possible initial conditions of physical systems (e.g. you are somehow prohibited from getting in the time machine that would take you back to kill your grandfather).  Deutsch uses his model to argue that, when quantum mechanics is taken into account, these unnatural constraints on initial states disappear.  Deutsch's fixed point theorem states that CTCs ``place no retrospective constraints on the state of a quantum system" \cite{Deutsch1991}.  That is to say, for any possible input state, there will be a paradox-free solution.

This is the result of a consistency condition implied by the quantum mechanical treatment of time-traveling carrier particles interacting with later versions of themselves.  If we let \(\left|\psi\right>\) be the initial state of the ``younger" version of the carrier particle, and let \(\hat{\rho}\) be the density operator of the ``older" version of the carrier particle, then the joint density operator of the two particles entering the region of interaction is \begin{equation}\left|\psi\right>\left<\psi\right|\otimes\hat{\rho}\end{equation} and the density operator of the two carrier particles after the interaction is \begin{equation}U(\left|\psi\right>\left<\psi\right|\otimes\hat{\rho})U^{\dagger}\end{equation} where \emph{U} is the interaction unitary.  The consistency condition requires that the density operator of the younger version of the carrier particle as it leaves the region of interaction is the same as that of the older version as it enters the region of interaction.

 This makes intuitive sense, because it is the interaction that causes the earlier version of the carrier particle to become the later version.  When translated via a denotationally trivial transformation to a network in which the chronology-violating behavior is localized to a single particle on a CTC that interacts with a chronology-respecting (CR) carrier particle, the consistency condition for the CTC system is \begin{equation} \rho_{\textrm{\scriptsize{CTC}}}=\textrm{Tr}_{\textrm{\scriptsize{CR}}} [U(\left|\psi\right>\left<\psi\right|\otimes\rho_{\textrm{\scriptsize{CTC}}})U^{\dagger}]. \end{equation} This requirement says that, after tracing out the CR qubit, the density operator of the system on the CTC \emph{after} the interaction is the same as it was \emph{before} the interaction.  That is to say, after the interaction, the carrier particle on the CTC enters the ``future mouth" of the CTC, and exits the ``past mouth" of the CTC \emph{before} the interaction.  The state of the particle that comes out of the past mouth must be the same as the system that enters the future mouth.\footnote{Because the procedure involves taking the partial trace of the system, and requiring consistency for only the state of the system bound to the CTC, any entanglement with systems in CR region is broken when the CTC-bound quibit exits the past mouth of the CTC.}  Furthermore, \(\rho_{\textrm{\scriptsize CTC}}\) depends on \(\left|\psi\right>\), so the input state on the causality-respecting carrier particle has an effect on the state of the particle it will interact with.
 
 The output of the circuit (i.e. the final state of the CR qubits) depends on the input of system \(\left|\psi\right>\) and \(\rho_{\textrm{\scriptsize CTC}}\). And, as we see in the previous equation, \(\rho_{\textrm{\scriptsize CTC}}\) itself depends on \(\left|\psi\right>\). Therefore, the evolution of the CR qubit is nonlinear with respect to the input \(\left|\psi\right>\). \begin{equation}\rho_{\textrm{\scriptsize{output}}}=\textrm{Tr}_{\textrm{\scriptsize CTC}}[U(\left|\psi\right>\left<\psi\right|\otimes\rho_{\textrm{\scriptsize CTC}})U^{\dagger}].\end{equation}

Whatever the physical situation is, its information flow can be redescribed in a form that has the following features: There are a finite number of qubits bound to a CTC. These interact via unitaries with a finite number of qubits that follow an ordinary chronology--respecting trajectory. The CR qubits are measured after their interaction with the CTC qubits, and their state is the final state of the system. In the region of interaction, the CTC qubits behave according to ordinary quantum mechanics, and interact with the CR qubits via unitary interactions. The CTC qubits do not evolve in any way while traveling back along the CTC. The nonlinearity of the systemÕs overall evolution is entirely due to the consistency conditionÕs nonlinearity. 

This means that the closed information loops of chronology violation can be isolated into localized regions of spacetime. The effects that can be generated by interaction with the CTC can range over all of space, of course. But they must be the result of entanglement, or prior causal interaction, with systems in the region of interaction with the chronology violating qubits.

In light of the model's reliance on this nonlinear consistency condition, Deutsch's claim that CTCs, when properly understood, place no constraints on the possible states of the quantum system may be stronger than is warranted.  While it is true that, unlike the classical analysis of time travel paradoxes, his model places no constraints on the input state of the causality-respecting system, it \emph{does} constrain the possible states of the system confined to the CTC.

While Deutsch's solution seems more intuitively plausible than the constraint on initial conditions that prevents the occurrence of classical time travel paradoxes, it is nonetheless puzzling.  In the classical case, it is somehow forbidden that I get in the time machine that will take me back to kill my grandfather.  There isn't necessarily any obvious causal mechanism that prevents me.  It is simply impossible, to avoid paradox, that I ever actually carry out my mission.  This constraint is often described as \emph{superdeterministic}, since it is something above and beyond simple determinism that rules out the possibility of me getting into the time machine. David Lewis's influential formulation of the classical consistency condition from his \cite{Lewis1976} alleviates some of this tension by redescribing the time travel narrative as a single, self-consistent history. The drawback of this approach is that it seriously undermines the notion that the time traveler has free will.

Deutsch characterizes his problem with the classical solutions as stemming from the fact that they violate what he calls the \emph{principle of autonomy}. \begin{quote} According to this principle, it is possible to create in our immediate environment any configuration of matter the laws of physics permit locally, without reference to what the rest of the universe may be doing. \cite{DeutschAndLockwood1994}\end{quote} He claims that classical solutions to the Grandfather Paradox, which impose global consistency, violate this principle. \begin{quote}Under this principle, the world outside the laboratory can physically constrain our actions inside, even if everything we do is consistent, locally, with the laws of physics. Ordinarily we are unaware of this constraint, because the autonomy and consistency principles never come into conflict. But classically, in the presence of CTCs, they do. \cite{DeutschAndLockwood1994} \end{quote}

Although Deutsch's reference to the principle of autonomy makes it seem as though his problem with the classical solutions is that they always rule out certain initial conditions of an experiment involving a CTC, this can't be quite right. For most initial setup conditions, including those trajectories that seem to entail a Grandfather Paradox, there is a fixed--point solution which shows that there is a self-consistent sequence of events that will avoid the paradoxical outcome. Often these solutions involve a self-interaction that changes the state of the younger system heading toward the future mouth of the CTC, such that when it exists the past mouth, it is no longer on a trajectory that will prevent it from time traveling. Novikov explains this class of solutions in terms of billiard ball entering the future mouth of a CTC on a trajectory that will lead to a collision with itself in the past, preventing it from entering the future mouth.\begin{quote} If we take into account the collision from the very beginning, then the collision is very weak, just a slight touch between the two balls that nudges the younger ball only slightly. The younger ball then moves along a trajectory slightly different from our expectation, but still enters mouth \(B\). It reappears from mouth \(A\) in the past and continues along its motion, still on a trajectory that differs only slightly from the trajectory it would have traveled on had it not suffered a collision. The result of the slight difference in trajectory is that the collision with the younger version of itself is not a strong collision, but rather a weak collision, a glancing blow. Therefore we have a consistent solution. \cite{Novikov2002}\end{quote}

There are some situations with which Deutsch is concerned that he claims do not have a classical fixed point. For example, the classical bit in the system depicted in Figure 1. After exiting the past mouth of the CTC, the system encounters a NOT gate, which flips its state. It then enters the future mouth, and repeats the process. Classically, this system would oscillate between the two values allowed.\footnote{Deutsch admits that discreteness in the classical domain is an approximation, so perhaps an argument could be made that the fixed point of such a system would involve a failure of the gate to operate properly. But assuming everything works as advertised, there is no classical fixed point.}

\begin{figure}[h]
\centering
\caption{\footnotesize{A classical bit bound to a CTC with a \emph{NOT} gate. In this example, there is no classical fixed point solution.}}
\includegraphics[width=.85\textwidth]{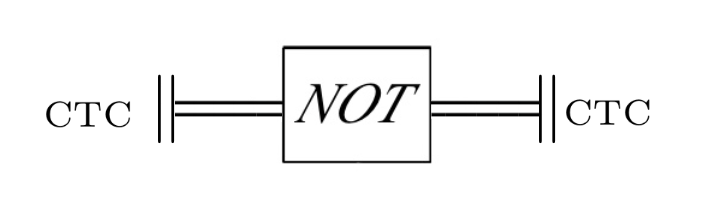}
\end{figure}

In the quantum version of this circuit, however, there is a unique fixed point. If the qubit is in the mixed state \begin{equation} \frac{1}{2}\left( \left|0\right>\left<0\right| + \left|1\right>\left<1\right|  \right) \end{equation} the consistency is satisfied.\footnote{As is true in the classical case, most fixed--point solutions in the D-CTC model are not unique. However, the information flow in this circuit is at the heart of the Grandfather Paradox, and is therefore especially important for Deutsch's goal to show that his model can solve the paradoxes of time travel. The BHW circuit, with which we will be concerned in the following, also has a unique fixed--point solution.}

The classical fixed point solutions show that in most cases, you are not constrained in terms of the initial setup conditions of your experiment. For most input states, there is at least one self-consistent sequence of events that can follow. I believe Deutsch is concerned with the fact that we are not free, even in light of the classical fixed--point solutions, to send a system in any state we choose \emph{into the CTC}. The classical fixed--point solutions involve a self-interaction before the input system enters the future mouth of the CTC. This, I believe, is at the heart of Deutsch's discomfort with the classical solutions to the paradoxes of time travel.

In Deutsch's model, this tension is seemingly resolved. A system in any state is allowed to enter the CTC---the time traveler could enter the time machine with any intentions whatsoever. Consistency is guaranteed by the state of the system confined to the CTC.  This doesn't offend the intuitions as badly as the classical case, because we can imagine the following pseudotime narrative:  The causality-respecting qubit begins its journey in some initial state, then encounters and interacts with CTC qubit, precipitating a change of state of both of them.  The CTC qubit in its new state then travels back in time to again interact with the causality-respecting qubit (in its initial state), and the interaction again changes the state of the CTC qubit.  Over infinite iterations of this process, the CTC qubit converges on some particular state, like the rotation of a top stabilizes after some initial wobbling. The CR qubit \emph{causes} the CTC qubit to be in the right state.

The puzzle arises, though, when we note that the CTC qubit must \emph{always} have been in this stable state.  There are no previous interactions with the CR qubit to force it to evolve over time into the right state.  So although Deutsch's model has avoided the superdeterminism of the traditional time travel paradoxes, which constrained the initial states of the CR system, it seems to have introduced significant kinematic constraints in another place. Something like Novikov's classical consistency condition must still be at play. That is to say, there must be a deeper metaphysical justification (i.e.\ the impossibility of a self-contradictory history) which is behind Deutcsh's quantum condition. And as we'll see in Section 5.2, Deutsch seemingly has something like this in mind. 

Deutsch's solution relies on allowing the system confined to the CTC to exist in a mixed state. Another puzzle arises when attempting to interpret what this can mean if, for example, there is only one CTC-bound qubit. As we'll see, the interpretation Deutsch gives of these mixed states relies on deep metaphysical commitments. The justification both for the existence of the consistency condition and for the existence of mixed states on the CTC comes from prior philosophical considerations Deutsch presupposes in the construction of his model.

Deutsch's analysis of the physical effects of chronology-violating regions of spacetime in terms of quantum computational circuits and the consistency condition has been very influential in the study of quantum information, and has led to many interesting insights about the nature of the quantum world.  One particularly interesting result is due to Brun, Harrington and Wilde.  In what follows, I will discuss their work, the debate surrounding their central claim, and further implications of their argument.


\section{The BHW Circuit}

Brun, Harrington, and Wilde, in their 2009 paper \cite{BrunEtAl2009}, describe a procedure for using CTC-assisted quantum computational circuits to distinguish between non-orthogonal states of a qubit. Ordinary linear quantum mechanics does not allow for such a discrimination to take place. It is only in the presence of certain kinds of nonlinear evolutions that nonorthogonal states can be projected onto orthogonal bases, making them reliably experimentally distinguishable. In this section, I will describe the protocol for distinguishing between the linearly dependent BB84 states \(\left|0\right>\), \(\left|1\right>\), \(\left|+\right>\) and \(\left|-\right>\), where \(\left|\pm\right>=\frac{1}{\sqrt{2}}(\left|0\right>\pm\left|1\right>)\).

\subsection{Details of the BHW Circuit}

The authors begin by detailing a protocol for distinguishing between two non-orthogonal states.  The setup involves two qubits: system \emph{A} in the unknown initial state \(\left|\psi\right>\) (either \(\left|0\right>\) or \(\left|-\right>\)), and system \emph{B}, a qubit in some state \(\rho_{\textrm{\scriptsize CTC}}\) on a CTC.  The procedure is simple: (1) perform a \(SWAP\) of systems \emph{A} and \emph{B}, (2) perform a controlled-Hadamard transformation with system \emph{A} as the control and system \emph{B} as the target, and (3) measure system \emph{A} in the computational basis.  A measurement of system \emph{A} that yields the output \(\left|0\right>\) means that the input state is \(\left|\psi\right>=\left|0\right>\).  A measurement of system \emph{A} that yields the output \(\left|1\right>\) means that the input state is \(\left|\psi\right>=\left|-\right>\).

This result obtains because of Deutsch's consistency condition.  Whatever state system \emph{B} is in when it enters the future mouth of the CTC must be the same state that comes out of the past mouth of the CTC.  That is, steps (1) and (2) must have no net effect on system \emph{B}.  The density matrix of the system on the CTC (system \(B\)) depends on the the input state of system \(A\), as shown in equation 3. The final state of the CR qubits depend on the input state of system \(A\) and on \(\rho_{\textrm{\scriptsize CTC}}\), as shown in equation 4. Since the only two possible input states are \(\left|\psi\right>=\left|0\right>\) and \(\left|\psi\right>=\left|-\right>\), the consistency condition requires that the only possible initial states of system \emph{B} are \(\rho_{\textrm{\scriptsize CTC}}=\left|0\right>\left<0\right|\) and \(\rho_{\textrm{\scriptsize CTC}}=\left|1\right>\left<1\right|\).

\begin{figure}[h]
\centering
\caption{\footnotesize{BHW circuit for distinguishing between \(\left|\psi\right>=\left|0\right>\) and \(\left|\psi\right>=\left|-\right>\) (from Brun et al.\ 2009).}}
\includegraphics[width=.85\textwidth]{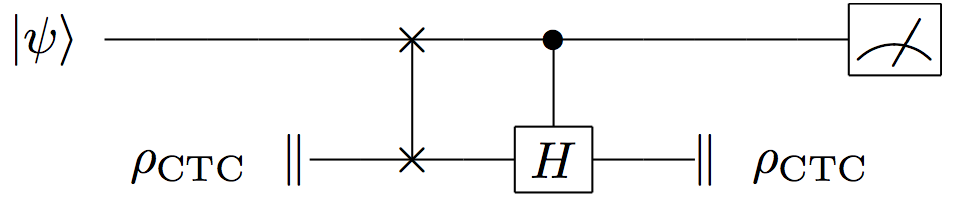}
\end{figure}

Consider the situation where the input state of system \emph{A} is \(\left|\psi\right>=\left|0\right>\).  If the initial state of system \emph{B} is \(\rho_{\textrm{\scriptsize CTC}}=\left|1\right>\left<1\right|\), then the effect of the first gate (\(SWAP\)) would be to transform system \emph{A} into the state state \(\left|1\right>\) and system \emph{B} into the state \(\left|0\right>\).  Since system \emph{A} is in the state \(\left|1\right>\), the action of the second gate (controlled-Hadamard with \emph{A} as the control and \emph{B} as the target) would transform system \emph{B} into the state \(\left|+\right>\).  Since the consistency condition requires that the state of \emph{B} after the action of the two gates is the same as the state of \emph{B} before the action of the two gates, it is clear that \(\rho_{\textrm{\scriptsize CTC}}=\left|1\right>\left<1\right|\) is not an allowed initial state of system \emph{B} when \(\left|\psi\right>=\left|0\right>\).

However, if the initial state of system \emph{B} were \(\rho_{\textrm{\scriptsize CTC}}=\left|0\right>\left<0\right|\), then after the action of the first gate (\(SWAP\)), system \emph{A} would be in state \(\left|0\right>\) and system \emph{B} would be in state \(\left|0\right>\).  The second gate (controlled-Hadamard) would not be activated since the control qubit is in state \(\left|0\right>\), so the consistency condition for system \emph{B} holds.  The measurement of system \emph{A} would yield a result of \(\left|0\right>\), which indicates that the initial input state was \(\left|\psi\right>=\left|0\right>\).

Now consider the case where system \emph{A} is initially in the state \(\left|\psi\right>=\left|-\right>\).  If the initial state of system \emph{B} is \(\rho_{\textrm{\scriptsize CTC}}=\left|0\right>\left<0\right|\), then after the action of the first gate (\(SWAP\)), system \emph{A} would be in the state \(\left|0\right>\) and system \emph{B} would be in the state \(\left|-\right>\).  Since \emph{A} is the control qubit for the second gate (controlled-Hadamard), it would not be activated and system \emph{B} would pass through unchanged.  It would therefore enter the future mouth of the CTC in the state \(\left|-\right>\), violating the consistency condition.

However, if system \emph{B} had initially been in the state \(\left|1\right>\), after the first gate, system \emph{A} would be in the state \(\left|1\right>\) and system \emph{B} would be in the state \(\left|-\right>\).  The control qubit would activate the controlled-Hadamard gate, and system \emph{B} would be transformed into the state \(\left|1\right>\), which is consistent with its original state.  The measurement on system \emph{A} will yield a result of \(\left|1\right>\), which indicates that the input was initially \(\left|\psi\right>=\left|-\right>\).

Brun and his collaborators were able to scale this protocol up to allow for the discrimination between the four non-orthogonal BB84 states \(\left|0\right>\), \(\left|1\right>\), \(\left|+\right>\) and \(\left|-\right>\).  They achieve this by adding an ancillary chronology-respecting qubit in the state \(\left|0\right>\), using two CTC-bound qubits, performing two SWAPs and four controlled unitary transformations, and making two measurements.

\begin{figure}[h]
\centering
\caption{\footnotesize{BHW circuit for distinguishing the four BB84 states (from Brun et al.\ 2009).}}
\includegraphics[width=\textwidth]{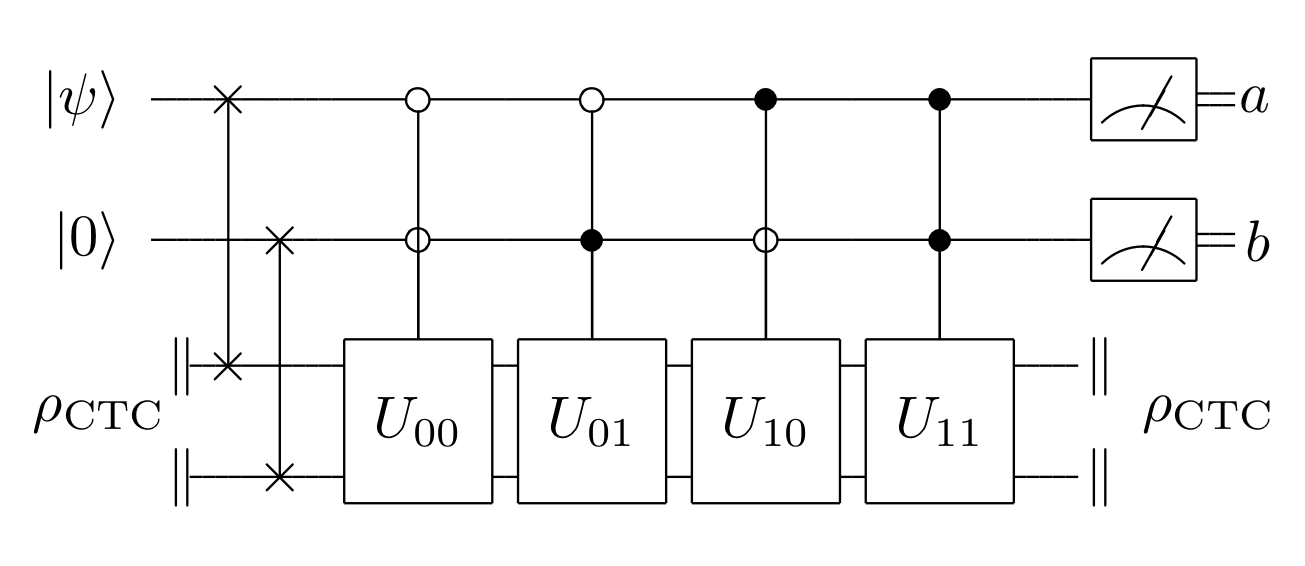}
\end{figure}The unitary transformations are as follows: \[U_{00}\equiv SWAP\] \[U_{01} \equiv X\otimes X\] \[U_{10} \equiv (X \otimes I) \circ (H \otimes I)\] \[U_{11}\equiv (X \otimes H) \circ (\textrm{SWAP})\]  The circuit performs the following map (\(\left|\psi 0\right> \rightarrow \left| ab\right>\)): \[\left|00\right> \rightarrow \left|00\right>\] \[\left|10\right> \rightarrow \left|01\right> \] \[\left|+0\right> \rightarrow \left|10 \right>\] \[\left|-0\right> \rightarrow \left|11\right>\] 

\subsection{Using BHW to Signal}

In Cavalcanti et al.'s \cite{CavalcantiEtAl2012}, the authors point out that the evolution of the quantum state through the BHW circuit which allows for the possibility of distinguishing the BB84 states is of the right kind to fit into a protocol for instantaneous signaling proposed by Gisin \cite{Gisin1990}.

Gisin's proposal involves two players, Alice and Bob, each sharing one half of a singlet pair.  Alice measures her particle either in the \emph{X} direction (yielding \(\left|1\right>\) or \(\left|0\right>\)) or the \emph{Z} direction (yielding \(\left|+\right>\) or \(\left|-\right>\)), forcing Bob's particle into the same state.  Bob then subjects his particle to a nonlinear evolution of a certain type that allows him to determine its state.  Gisin proposed a particular nonlinear Hamiltonian that would do the job, but Cavalcanti and Menicucci point out that the BHW circuit has the right features to fit into this framework.  

\begin{figure}[h]
\centering
\caption{\footnotesize{The method for using the BHW circuit in Gisin's instantaneous signaling device.  Alice measures first measures her particle along the \emph{X} or \emph{Z} axis.  Bob then uses the BHW circuit to determine what state his particle is in.}} 
\includegraphics[width=\textwidth]{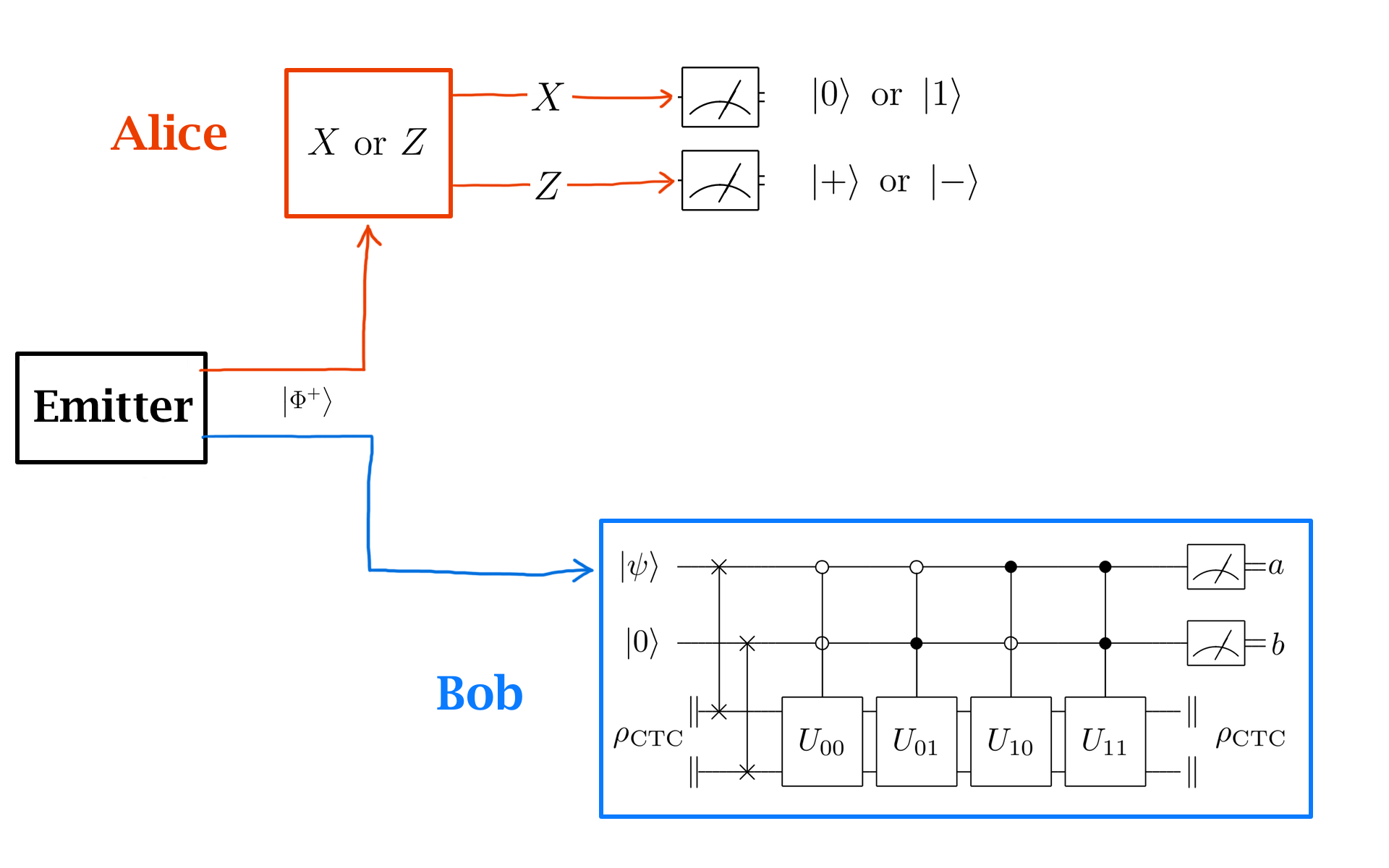}
\end{figure}

The BHW circuit will allow Bob to perfectly distinguish between all four states.  Therefore, if Alice wants to send a 1-bit message, she can choose either to measure in the \emph{X} direction (for ``yes") or the \emph{Z} direction (for ``no").  Bob, using the BHW circuit, can recover Alice's message, which is transmitted instantaneously.  That is, if the output of Bob's device is \(\left|10\right>\) or \(\left|11\right>\), he knows Alice measured her half of the singlet pair in the \emph{X} direction (intending the message to be ``yes"), and if his results are either \(\left|00\right>\) or \(\left|01\right>\), he knows she measured in the \emph{Z} direction (meaning ``no"). 

\subsection{The Bub-Stairs Consistency Condition}

In a recent paper \cite{BubStairs2014} Jeffrey Bub and Allen Stairs propose a consistency condition to solve one of the outstanding conceptual problems with Nonlocal Signaling. Their condition solves some potential ambiguity associated with the possibility of signaling.

The issue that the consistency condition is designed to solve arises because of the fact that the nature of Nonlocal Signaling allows for cause/effect to happen at spacelike separation. Alice's choice of measurement causes Bob to get the result he does faster than a light signal would have been able to traverse the distance. Since the event of Alice's input and the event of Bob's output are at spacelike separation, observers in different frames will disagree about which event comes before the other.That is, for some observers, Bob will measure his particle before Alice measures hers. In those frames, Alice and Bob's shared Bell State will not have been disentangled by Alice's measurement, and therefore Bob will input a particle in the state \(I/2\) into the BHW circuit, which will yield any of the four possible outcomes with equal probability, meaning that he has a 1/2 probability, in that frame, of getting an output that corresponds to the wrong input for Alice. In frames where Alice measures first, her choice determines Bob's output by disentangling their shared Bell State, leading to Bob measuring a particle in a definite state with the BHW circuit. In frames where Bob measures first, he inputs his still-entangled particle into the BHW circuit, yielding each of the four possible outputs with equal probability, regardless of the input Alice later chooses.

To protect against this sort of problem, they introduce a simple and elegant new consistency condition. It consists of the conjunction of the following two claims: \begin{enumerate} \item[(C1)] Observers in differently moving reference frames agree on which events occur, even if they disagree about the order of events.
\item[(C2)] If an event has zero probability in any frame of reference, it does not occur. \end{enumerate}

C1 ensures the two observers would agree about the outcomes of the two measurements (namely that the output Bob gets corresponds to the input Alice makes, regardless of who goes first). C2 ensures that the contradiction will never arise, since according to one observer, the probability of Bob getting the outcome that is inconsistent with Alice's input is 0.

While the consistency condition seems unobjectionable, I'll argue in Section 5.3 that the conclusions Bub and Stairs attempt to draw from it are more problematic.

\section{Signaling and Relativity}

It's not uncommon at this point in the discussion for the argument to be made that relativity prohibits signaling. Consistency with special relativity is taken to rule out the possibility of superluminal causal influence, or superluminal information transfer, both of which feature in the BHW signaling protocol. However, it has been argued convincingly in a number of places that the sending of superluminal signals is itself inconsistent with relativity. For example, Maudlin showed in \cite{Maudlin2011} that there are hypersurfaces of superluminal signal reception that are Lorentz invariant. Furthermore, he argues that even in the case of a locus of reception that falls along the simultaneity surface defined by the state of motion of the emission source of the signal, there isn't a direct contradiction with relativistic constraints. \begin{quote} On the other hand, Relativity \emph{per se} in no way constraints case-1 signals. It is consistent even with case-1 superluminal signals which propagate instantaneously (i.e.\ along a flat space-like hyperplane). But the hyperplane must be in part determined by the state of the emitter, or of some other matter. \cite{Maudlin2011}  \end{quote} What is problematic from a relativistic point of view is the fact that information can be sent to the past by chaining together superluminal signals, and in so doing, we apparently give rise to the possibility of contradictions (like the Grandfather Paradox).\footnote{The simple objection that superluminal signaling would violate relativity because the information is ``traveling faster than light" can be answered by pointing out that there is nothing about exploiting quantum nonlocality to sent usable information that requires that a carrier of information physically traverses the space between the communicating parties. Quantum teleportation arguably involves sending quantum information (in the form of the unknown state \(\left|\psi\right>\)) faster than light from Alice to Bob, though in order for Bob to use it, a two--bit classical message  must be sent at subluminal speeds. See Timpson's discussion of teleportation in \cite{Timpson2013}.} For example, Timpson says \begin{quote} The constraint is that superluminal signalling is ruled out on pain of temporal loop paradoxes. What this means is that no physical process is permissible that would allow a signal to be sent superluminally and thus allow information to be transmitted superluminally. \cite{Timpson2013} \end{quote} This view is typical (see e.g.\ \cite{Maudlin2011} and \cite{Norton2013}).

It would be possible to prevent this outcome by defining a privileged Lorentz frame in which the signals could travel their maximum (potentially infinite) speed, and in all other frames, it travels at lower speeds. However, there is some disagreement about whether this solution would itself be in conflict with relativity. Maudlin says that a situation in which there is a privileged frame, a ``fundamental relativity principle would be violated" \cite{Maudlin2011}. Nicolas Gisin disagrees, saying it does not violate relativity. \begin{quote}However, the assumption of a universal privileged reference frame with respect to which a faster than light influence can be defined, is not in contradiction with relativity. \cite{Gisin2012} \end{quote} In this context, Gisin develops another series of models of superluminal causal influence, arguing that the existence of such an influence can always be exploited for superluminal signaling.

Another possible solution to the problem of contradictions following from the ability to send information to the past is to impose some philosophically--motivated condition to rule out inconsistent histories. This is exactly the function of consistency conditions in discussions of chronology violation. Classical discussions of CTCs include versions of the Novikov condition. And Deutsch's consistency condition is a quantum analogue. Maudlin claims that any condition that could rule out temporal paradoxes in the context of superluminal signaling would most likely be in contradiction to relativity. \begin{quote} So the choice we have is not between superluminal signals and Relativity but between superluminal signals which allow for loops and Relativity. If we can argue that loops are impossible, then accepting the signals means positing extra structure to space-time to forbid them, structure that looks suspiciously like an absolute notion of simultaneity.  \cite{Maudlin2011} \end{quote} Therefore, any attempt of this kind would be in conflict with relativity. Whether a CTC consistency condition is equivalent to a privileged reference frame is an interesting question, but is beyond the scope of the present work.

Here it will suffice to say that the origin of the superluminal signaling in this debate is from the assumption that CTCs behave in such a way that Deutsch's consistency condition holds. It is the consistency condition that induces the nonlinear evolution that leads to the ability to distinguish the BB84 states. However, there is the question of whether the D-CTC consistency condition should apply to information sent to the past in this way. Deutsch's model is motivated by the possible existence of CTCs as understood from a relativistic point of view. He mentions the consistency of wormholes with general relativity in the setup of his argument. But he also explicitly abstracts away from these spacetime structures in developing his model. He considers CTCs to be characterized by a closed path of information. If there is no closed path for information, then there is a denotationally trivial transformation that can eliminate all negative time delays. A spacetime that contains any effect of a closed loop of information is subject to his analysis. \begin{quote}   Negative delay components in the model play the role of time machines, which I define in general as objects in which some phenomenon characteristic only of chronology violation can reliably be observed. [...] The basic method of this paper is to regard computations as representative physical processes---representing the behavior of general physical systems under the unfamiliar circumstances of chronology violation. \cite{Deutsch1991}\end{quote} According to Deutsch, any physical system in which there is a closed loop of information can be represented as a CTC-supplemented computational circuit. Though he never explicitly addresses this issue, from what has been said, it seems clear that this should also apply to cases where the closed loop of information is made possible by the prior existence of a CTC, and the loop exists outside of what was originally taken to be the region of interaction. [Um, can you transform a simple signaling setup into the standard form?]

So it seems that objections to signaling in the context of this debate cannot rely purely on the claim that it is inconsistent with relativity. But the debate surrounding the BHW circuit and related results includes people making this  claim (see e.g.\ \cite{BennettEtAl2009} and \cite{CavalcantiAndMenicucci2010}). I will argue in the next section that this reluctance to allow superluminal signaling is in part motivated by the status of the No-Signaling principle in quantum information theory. I will argue that this way of thinking about quantum mechanics is in tension with the fundamental commitments that underwrite Deutsch's D-CTC model.

\section{Quantum Information-Theoretic Motivations}

\subsection{Why Maintain No Signaling?}

I've argued that the relativistic justification for the No-Signaling Principle doesn't stand up to scrutiny. I will argue that the reason for the reticence to give up the No-Signaling Principle has to do with its status in the QIT. In particular, it is one of the most promising principles used in the reconstructions of quantum theory from the space of generalized probabilistic theories. This research project considers the space of all possible theories formulated in terms of their information processing capabilities and their allowed correlations between events. \begin{quote} For any theory, whether it applies to Nature or not, one can consider the information processing possibilities of this theory, the differences from those of classical or quantum theory, and attempt to trace these possibilities back to the fundamental features of the theory. \cite{Barrett2007} \end{quote} Principles are introduced, which partition the theories in that generalized theory space. The hope is that a small number of physically plausible principles can be identified which will pick out exactly those correlations allowed by quantum theory. On this view, these principles would give us an answer to the question of why our world allows for the quantum correlations to obtain, and not others.

No Signaling is one of the core principles at the heart of this approach. If the privileged status of the No Signaling principle were to be undermined, that would in turn undermine the status of other principles promising for this project, which live or die with No Signaling. One such example is the principle of ``information causality", which generalizes the No Signaling principle in the following way. If Alice sends Bob \(m\) classical bits, the most classical information Bob can extract from that message is \(m\) bits. This reduces to the No Signaling principle in the case where \(m=0\) \cite{pawlowski2009}. This is consistent with the teleportation protocol because, even though Alice can send Bob a quantum state that would take a potentially infinite amount of classical information to perfectly specify, Bob cannot extract more than the 2 classical bits Alice sent to him. Having Alice's input state \(\left|\psi\right>\) in hand does not give Bob any more information.

This approach is closely associated with the principle-based conception of physical theories. This is a conception in which the fundamental formulation of a physical theory is in terms of principled restrictions on the kinematical level. These principles never need to be justified by ontological or dynamical considerations. An empirically equivalent theory theory formulated in terms of dynamics and ontology is taken on this approach to represent a less fundamental formulation.

The special theory of relativity is taken to be the paradigm example of this kind of theory. Just as the Principle of Relativity and the Light Postulate pick out Minkowski spacetime as the space of events in SR, and constrain the structure of events in spacetime, the information-theoretic interpretation of quantum theory take there to be principles that define a space of events for quantum theory, and that space to constrain the structure of those events.

\begin{quote} In the case of quantum mechanics, these principles are information-theoretic and include a `no signaling' principle and a `no cloning' principle. The structure of Hilbert space imposes kinematic (i.e.\ pre-dymanic) objective probabilistic constraints on events to which a quantum dynamics of matter and fields is required to conform, through its symmetries [...].   \cite{TwoDog} \end{quote} And, as with relativity, they hold that there is no deeper explanation of the structure of events than that they are subject to the constraints embodied in the principles.

\begin{quote} There is no deeper explanation for the quantum phenomena of interference and entanglement than that provided by the structure of Hilbert space, just as there is no deeper explanation for the relativistic phenomena of Lorentz contraction and time dilation than that provided by the structure of Minkowski spacetime. \cite{TwoDog}  \end{quote}

In most cases these two ways of formulating a theory (principle and constructive) don't come into any kind of conflict. But in the case of the nonlinear extensions of quantum theory, the principle version of the theory makes different predictions than the constructive version. Following the dynamics of the systems under consideration leads us to conclude that signaling is effected in the BHW circuit. But this is in explicit conflict with the No-Signaling Principle.

Proponents of the principle theoretic interpretation of quantum theory, like Bub and Pitowsky in \cite{TwoDog}, argue that there is a strict distinction between principle and constructive conceptions of physical theories. Either a theory is fundamentally formulated in terms of principles, or it is constructed from an ontology and dynamics. From the principle conception every explanation must reduce to the principles. And from the constructive conception, all principles must be explainable in terms of the ontology and dynamics.

\subsection{Deutsch's Metaphysics}

The core claim of this argument is that Deutsch's model requires a commitment to a deeper metaphysical picture than the principle--theoretic approach can support. I will argue in this section that Deutsch relies on a metaphysical commitment that goes beyond the standard Everettt interpretation, and it plays an ineliminable role in the predictions of the D-CTC model.

Deutsch frequently makes reference to the ``multiverse'' as the spacetime in which the events that solve the paradoxes of time travel occur. As an Everettian, he is committed to the existence of the branching structure that gives rise to the existence of many worlds. But there is additional structure needed to make the D-CTC model operate in the way he claims it does. For details, see \cite{Dunlap2015b} but briefly, he is committed to the existence of the Many--Worlds Multiverse (MWM), and an additional structure I call the Mixed-State Multiverse (MSM). 

The most obvious interpretation of the multiverse of which Deutsch is making use to ground the D-CTC model is the MWM. But there is an immediate problem with this interpretation of Deutsch's model. Imagine a time traveler traveling from \(t = 2\) back in time to \(t = 1\). She cannot be traveling into her own past, because her presence there would change it, leading to a different future evolution of the wavefunction, undermining the existence of the branch from which she came. She needs to travel to an already existent branch with an identical copy of herself at \(t = 1\). The problem is, according to the Everett interpretation, there would be no such branch. Since the state of the world at \(t = 1\) in the time traveler's actual past is, by stipulation, identical to the state of the world at \(t = 1\) in the universe into which she is traveling, there would never have been a branching event that had created multiple copies of the world. The existence of this destination world is not consistent with the branching structure of the standard Everett interpretation. Deutsch cannot be relying on the structure of MWM for his solution to the paradoxes of time travel.

The existence of the MSM is necessary for solutions of the following kind to obtain:\begin{quote} In all universes the observer approaches the chronology-violating region on a trajectory which would go back in time. But only in half of them does the observer remain on that trajectory, because in half the universes there is an encounter with an older version of the observer after which the younger version changes course and does not go back in time. After that, both versions live on into the unambiguous future. \cite{Deutsch1991}\end{quote} It is the parallel worlds of the MSM that are connected by CTCs. This structure is not implicit in MWI itself, and is added by Deutsch as a solution to the paradoxes of time travel.\footnote{There is a related result that suggests that Deutsch's solution to a simple Grandfather Paradox circuit would not hold if he weren't explicitly relying on the MSM structure. See \cite{DunlapDissertation}.}

Since Deutsch's metaphysics plays such a central role in the formulation of his CTC model, it cannot be ignored. It is therefore problematic for quantum information theorists to adopt this model for analysis in a principle-theoretic context. Despite their claim to be neutral to questions of interpretation, the metaphysics is playing a fundamental role in the example.

As a consequence, the ease of fit between principle--theoretic considerations that reify the No Signaling principle in a context where the fundamentality of underlying metaphysical considerations is denied, with a model that relies crucially on such commitments to yield the predicted behavior under consideration.

The D-CTC model requires a strong commitment to a particular ontology, and is therefore the product of a constructive version of quantum theory. On the constructive view, top--down principles are not considered to have any fundamental importance, and when they conflict with predictions based on the ontology and dynamics, they are discarded. The D-CTC model predicts signaling because of the commitments of the theoretical framework in which it was developed.

Can we embed the mathematical features of the D-CTC model in a principle--theoretic framework without making the same metaphysical commitments? As argued in \cite{Dunlap2015b}, denying the D-CTC model recourse to the explanatory resources provided by realism about the ontology of MSM will lead it to make different predications in simple cases, undermining its ability to solve even the Grandfather Paradox.

The D-CTC model includes elements of both the metaphysical approaches to quantum theory, and quantum information. Although these two approaches come into conflict, I believe Deutsch had a consistent view in mind. His analysis of the power of quantum computation, for example, is based on parallelism. He considers the increased capabilities of a quantum computer to be proof of the existence of the parallel worlds in which the many computations necessarily must take place. Although this strong metaphysical commitment to the existence of parallel worlds is a minority view among contemporary QIT researchers, it is not strictly in conflict with it. It is in conflict, however, with the purely operationalist interpretation of quantum mechanics that leads to the principle--theoretic conception of fundamental physical theories.

From the constructive perspective, ``principles" are corollaries of the constructive theory. And as such, if the dynamics that governs the behavior of the ontology changes (as in this nonlinear extension of QM), then the corollaries will potentially cease to hold. Deutsch holds to the constructive picture even with respect to the consistency condition and the evolutionary principle. These are explainable by virtue of his expanded ontology. In fact, without the particular features of his ontic picture constraining the possible outputs of the CTCs, the model won't work in precisely the way he says it does.

\subsection{Bub-Stairs Consistency}

Another minor conflict with Deutsch's approach arises in the Bub and Stairs paper, and is related to the point about nonlocality and relativity from above. Bub and Stairs argue that their consistency condition allows for a ``radio to the past", or a protocol for sending classical information back in time. They contend that the existence of this protocol opens the door to temporal paradox. As mentioned in Section 3.4.2, I believe that Deutsch's consistency condition wold apply to this classical information channel as well.

The evidence for this claim is comes from the fact that Deutsch, as an Everettian, would deny that there was any principled distinction between classical information and quantum information. Ultimately, classical information supervenes on quantum systems. In order to be consistent with his broader view on the interpretation of quantum theory, he must treat the classical domain and the quantum domain as subject to the same laws, particularly one as fundamental as a consistency condition.

In fact, in his paper, he explicitly states that he is conceiving of computation for the purposes of this argument as \begin{quote}a representative physical process---representing the behavior of general physical systems under the unfamiliar circumstances of chronology violation. \cite{Deutsch1991}\end{quote} He develops a standard form for a CTC-assisted quantum circuit for the purposes of defining his consistency condition in a simple way. But he says that any spacetime bounded network, which he uses to represent general physical systems, can be trivially transformed into a denotationally equivalent standard form, which localize any closed loop of information onto a CTC. \begin{quote} ...[T]he transformed version would be intuitively very different from the original one which might represent a time traveler, whereas the transformed version appears to represent an ordinary space traveler meeting a time traveler who spontaneously comes into existence as an identical twin of the space traveler, exists for a finite period of time on an ``eternal" loop, and then ceases to exist. \cite{Deutsch1991} \end{quote} It is clear that Deutsch takes these quantum-circuit representations to be completely general. Therefore, his consistency condition should apply to all physical systems.

Bub and Stairs consider the radio to the past protocol to be potentially paradoxical because they insist on a strict distinction between the classical domain and the quantum domain. They say that they see their consistency condition as allowing for a ``radio to the past'', which opens the door for the reemergence of the time travel paradoxes in the classical domain. This comes from the fact that they are implicitly taking on a Heisenberg (or operationalist) picture, which is characteristic of quantum information, but is rejected by the realist approaches to the interpretation of quantum theory. This is the same problem we saw above: the tenets of the quantum information-theoretic interpretation of quantum theory are doing work behind the scenes to justify the approach to the problem.

And finally, it should be noted that even in a purely classical context, there are analogues of Deutsch's consistency condition that are taken equally scientifically seriously (e.g. \cite{Lewis1976} and \cite{Novikov2002}). So even if they argue that Deutsch's consistency condition only applies to quantum information, there are consistency conditions in reserve ready to step in.

\section{Conclusion}

The considerable recent interest among the quantum information community in the D-CTC model has produced genuinely interesting results. The BHW circuit and its use in the Nonlocal Signaling protocol are considerable contributions to our understanding of how quantum systems behave in nonlinear extensions of quantum theory.

However, we must be sensitive to the fact that there are significant constraints on the generality of the D-CTC model. Its formulation presupposes significant metaphysical commitments, and is therefore applicable only in contexts where those metaphysical commitments are shared. Failing to recognize this feature of the model threatens to undermine its application. I argue that this problem is present in the debate in the quantum information literature, in particular in the attempts to impose the No-Signaling Principle on the framework in which the system is being analyzed.

Because of these underlying commitments, the D-CTC model serves as an important example for the divergence between the principle-theoretic approaches to quantum theory, and the more metaphysically robust constructive approaches. Deutsch himself is unambiguously an advocate of the latter, and the model is arguably incoherent on the former approach.



\end{document}